# Singularity-Free Collapse through Local Inflation [1]

Hunter Monroe
International Monetary Fund[2]

In common relativistic models of gravitational collapse, the collapsing star's interior experiences a bubble-like local inflation, allowing radii to diverge rather than converge toward a singularity. This proves a conjecture of Shatskiy [25]. If $\ell$ represents proper radial distance along a particle's trajectory in the collapsing star, then $dr/d\ell = 0$ and $d^2r/d\ell^2 > 0$ at the horizon, so the solution locally resembles the neck of a wormhole from the exterior to the interior. The implied limiting curvature is at the horizon not the Planck level.

## 1.   INTRODUCTION

In a spherically symmetric spacetime, it is a natural assumption that a radial coordinate $r$, where $r$ is scaled such that a sphere has surface area $4\pi r^2$, increases as one moves from the center outward. However, in non-Euclidean spacetime "there is no *a priori* reason to expect that the surface area $4\pi r^2$ and hence the radial coordinate $r$, will increase monotonically as one moves from the center of the star outward" (Misner *et al* [19]). Below, it is shown that in typical dynamic models of gravitational collapse, the interior experiences a local inflation, and $r$ reaches a local minimum at a horizon. As a result, the interior does contain a trapped surface and a singularity.[3]

## 2.   LOCAL INFLATION DURING GRAVITATIONAL COLLAPSE

Assume matter in a collapsing star is spherically symmetric and non-rotating. Define a particle horizon as follows. Suppose two initially collocated observers synchronize clocks, and let $t$ be the proper time of the first observer, and $\tau$ the proper time on the second observer's clock as read by the first observer (for instance through photons sent at regular intervals by the second observer to the first). Define the notional distance from the second to the first observer to be $t - \tau$. Then, if the first observer proceeds toward $r = +\infty$, $r = 2M$ is a particle horizon from the first observer's perspective, in that the notional distance from the second observer to the first increases without bound as the second observer approaches $r = 2M$.

---

[3] See Oppenheimer and Snyder [21], Penrose [22], Hawking and Penrose [11], and Hawking and Ellis [10].

Suppose two observers are at $r = kM$ for $k > 2$ and the second observer moves to $r = 0$ at non-relativistic speeds, sufficiently early in the collapse that time dilation is negligible. Then, the notional distance from $r = 0$ to $r = kM$ is initially $kM$. Notional distance will be an analytically useful measure of causal separation during collapse, because it increases continuously up to the point at which a notionally infinite distance between the observers prevents any communication from the first to the second.

In a two-dimensional diagram, let the vertical dimension represent the proper time $\tau$ of the second observer at an inertial frame at rest at the origin. Let the other dimension be the notional distance $\ell$ from the origin measured along a null line from the origin.[4] The diagram will also mark off on the $\ell$ coordinate the proper distance to a given Schwarzschild radial distance $r = kM$, showing the correspondence between the coordinate system of the second observer at the origin and the first observer at $r = kM$. On a null line from the origin with given $\tau$ coordinate, $r$ depends on $\ell$, and can be written as a function $r = r(\tau, \ell)$.

At $\tau = 0$, suppose the time dilation effects are negligible, so $\ell \approx r(0, \ell)$ and $dr/d\ell \approx 1$. From $\tau = 0$ up to some time $\tau = N$ while these effects remain negligible, the curve $r(\tau, \ell) = 2M$ will be vertical at $r = 2M$, and similarly for any $r = kM$. As the star collapses and time dilation intensifies, the notional distance from the origin to a given $r$ will increase.

---

[4] This coordinate is somewhat similar to an outgoing Eddington-Finkelstein coordinate (Misner *et al* Box 31.2).

Therefore, the curve $r(\tau, \ell) = 2M$ will bend to the right and $dr/d\ell$ decreases. Figure 1 shows the diagram at this point.

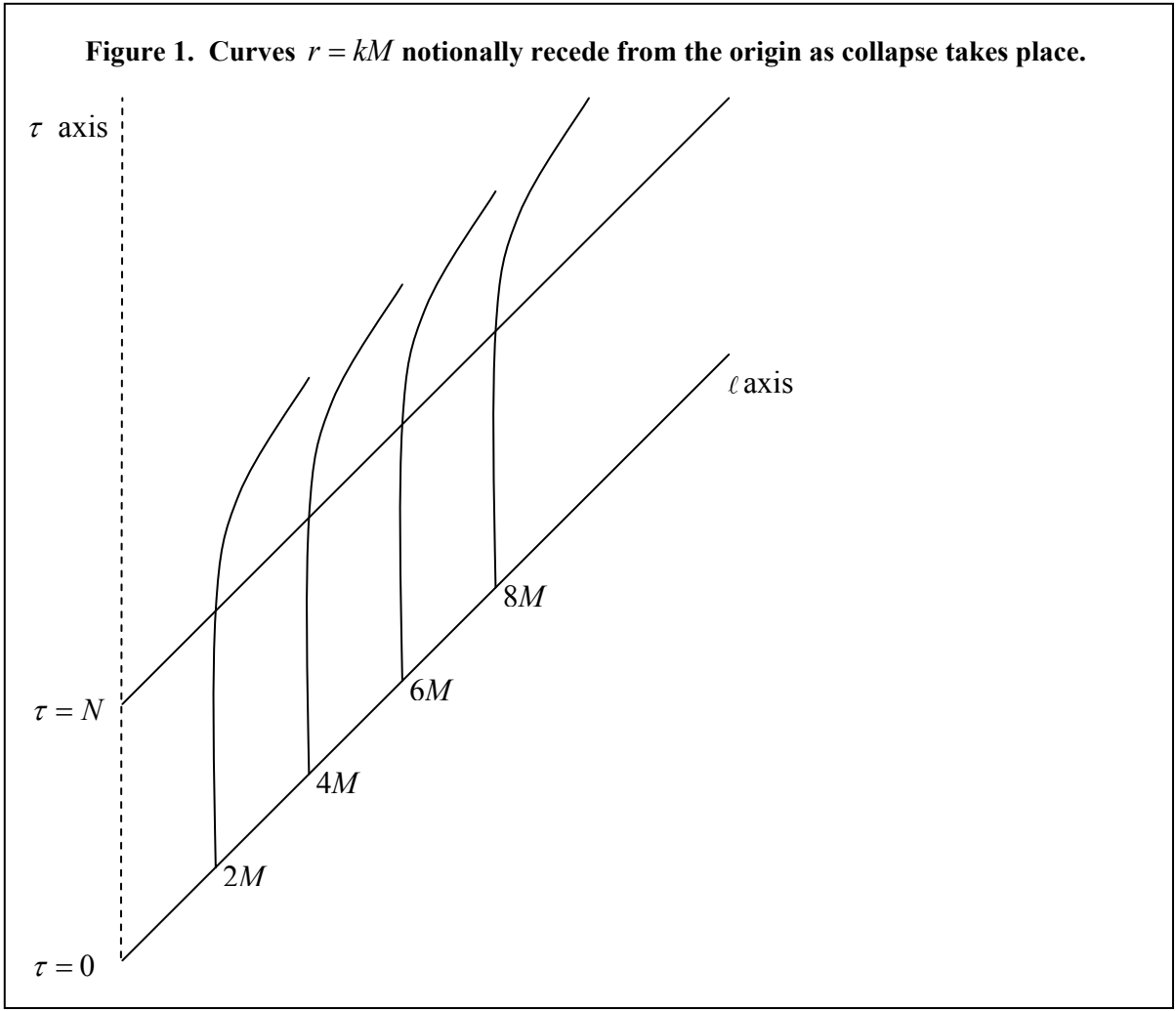

**Figure 1. Curves $r = kM$ notionally recede from the origin as collapse takes place.**

Assume that the function $r(\tau, \ell)$ has the following properties in the exterior, which can be readily verified in a given model of collapse, such as a model with uniform density and zero pressure (Misner *et al* §32.4). Broadly, these assumptions imply that as the star collapses, $r = 2M$ recedes smoothly to notional infinity by a finite time $\tau = E$.

**Property 1**. The function $r(\tau, \ell)$ is strictly increasing in $\ell$, because the speed of light is always positive.[5]

**Property 2**. After some time $\tau = N$, the function $r(\tau, \ell)$ is strictly increasing in $\tau$. That is, the process of collapse is monotonic after $\tau = N$, and is nearly Newtonian beforehand.

**Property 3**. The function $r(\tau, \ell)$ is continuous.

**Property 4**. A particle horizon emerges by a finite proper time $\tau$, that is,

$E = \inf\{\tau \,|\, \forall \ell : r(\tau, \ell) < 2M\} > N$.

Properties 1 and 4 imply $\forall \ell : r(E, \ell) < 2M$, that is, the null line at $E$ and the curve $r(\tau, \ell) = 2M$ converge asymptotically.[6] Furthermore, a point on the surface of the collapsing star will intersect the curve $r(\tau, \ell) = 2M$ at the moment that the curve and the null line intersect, which occurs at notional infinity. That is, the surface of the star does not intersect the null line at $\tau = E$ and the curve $r(\tau, \ell) = 2M$, but converges asymptotically to it. The properties above guarantee that a curve $r(\tau, \ell) = 2M - \varepsilon$ intersects the particle horizon for $\ell$ arbitrarily large, if $\varepsilon$ is sufficiently small.[7]

---

[5] See however the next footnote.

[6] This assumes matter is continuous rather than discrete. With discrete rather continuous matter, there would be a point in time at which the last particle of the collapsing star reaches $r$=2M. In that case, a modified Property 1 would call only for the function to be non-decreasing, and the null line at $E$ would intersect the curve $r(\tau, \ell) = 2M$ at the point $(E, \ell_{2M})$ and coincide for $\ell \geq \ell_{2M}$. In the case, the analysis below would change little, as the null line and the curve would nearly be tangent, given the large number of particles required for a collapsing star to form a black hole.

Figure 2 shows the diagram with this additional information. The four properties above imply that from the perspective of a point at the origin, the surface of the star at some point must stop collapsing toward the origin and begin receding to notional infinity, because the surface is converging asymptotically toward a curve $r(\tau, \ell) = 2M$ which is receding to notional infinity. Shatskiy [25] (Figure 4) conjectures a similar process in which the interior expands as collapse proceeds.

The last $\varepsilon$ (for $\varepsilon$ sufficiently small) of the collapsing star's mass to pass through the particle horizon emerges from the horizon at an arbitrarily large notional distance from the origin, implying that radii are diverging in the interior in a neighborhood of $r = 2M$.

**Theorem 1.**     In any model of gravitational collapse satisfying Properties 1-4, $dr/d\ell = 0$ and $d^2 r/d\ell^2 > 0$ at the particle horizon.

**Proof:** The theorem follows immediately from Properties 1-4. QED.

---

[7] This argument is valid if matter is continuous rather than discrete, or if the mass of a single particle is very small in relation to the mass of the black hole. See footnote 6.

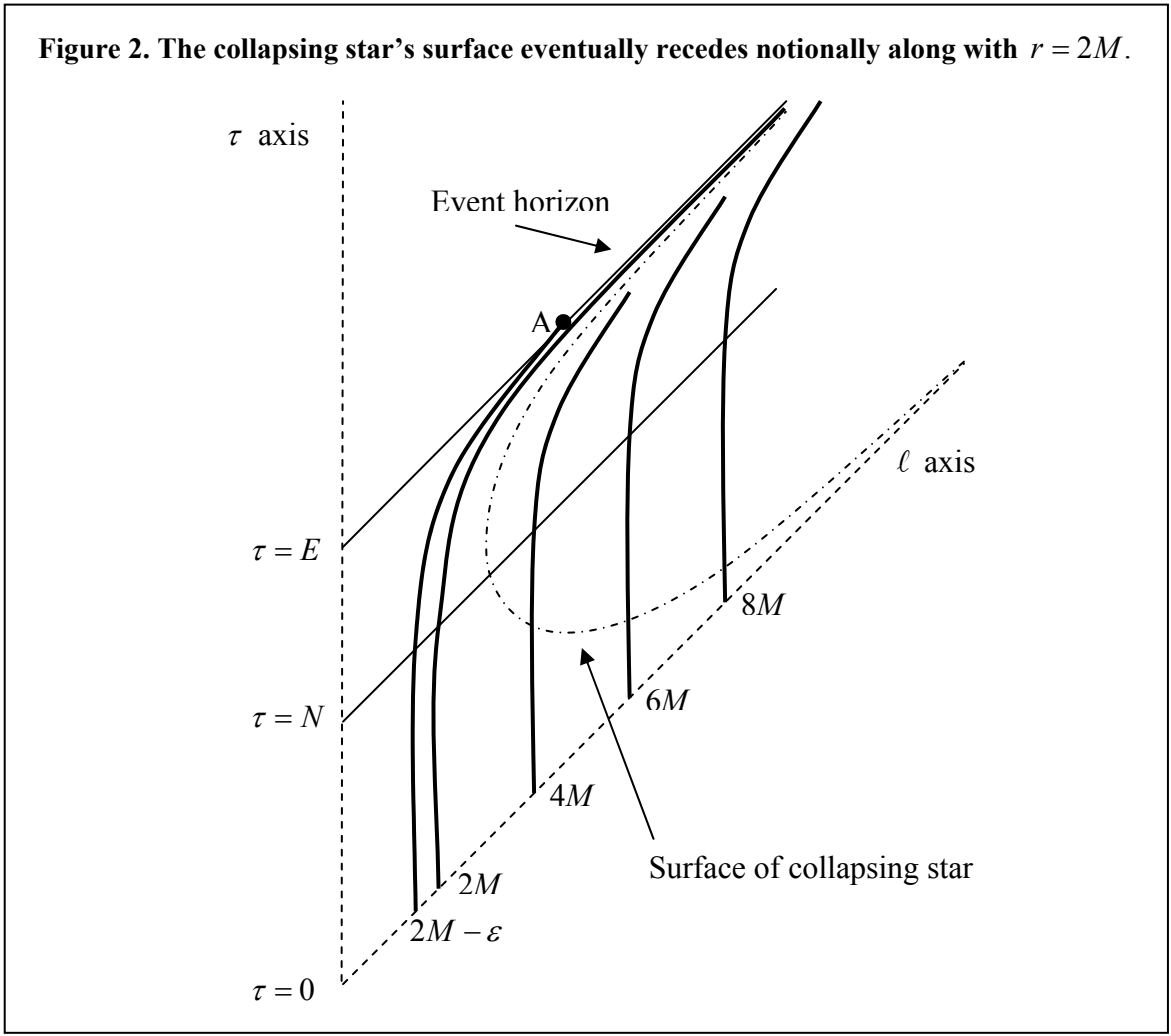

**Figure 2. The collapsing star's surface eventually recedes notionally along with** $r = 2M$.

The result above implies a limiting curvature which prevents escape velocity from exceeding the speed of light. Under the assumption of such a limiting curvature, Frolov *et al* [6][7] observe that black holes generate closed worlds. Easson and Brandenberger [4] note that the generation of universes from black hole interiors would solve the horizon, flatness, and structure formation problems without requiring a long period of inflation and also resolve the information loss and other paradoxes. Preskill [24] also concludes that baby universes are the most satisfying resolution to the information loss paradox.

The interior forms an expanding universe, in that particles traveling on diverging radii would become increasingly distant from each other (Armendáriz-Picón [1]). The rate of expansion (divergence) is determined exactly by the Einstein equation.

Hawking and Penrose [11] predict a singularity before the big bang, so at least one of this theorem's assumptions does not hold in an expanding interior. In particular, there is no time-reversed trapped surface; past-directed geodesics do not converge but cross the horizon at distinct points in spacetime.

### 3.    QUANTUM MECHANICS VS. GENERAL RELATIVITY

To the extent that Theorem 1 applies generically, general relativity does not predict its own breakdown at singularities, and does not become inconsistent with quantum mechanics in the neighborhood of singularities. This section briefly revisits the two other principal points of tension between the two theories: the Heisenberg uncertainty principle and non-locality.

### 3.1.    Uncertainty Principle

It is possible to derive an uncertainty principle in general relativity as follows, which qualitatively resembles the Heisenberg uncertainty principle.[8] The relativistic gravitational waves produced by a chaotic $n$-body system,[9] with $n$ equal to the number of massive particles in the observable universe, would cause each body to follow a deterministic but chaotic trajectory.[10] To an observer not able to predict the chaotic gravitational waves impinging on a given particle, the particle's location and velocity would appear stochastic.[11] A known example of this behavior is the Brownian motion of black holes at a galactic center, due to *n*-body behavior (Laun and Merritt [14]).

---

[8] Carati and Galgani's [3] review other quantum-like features arising in classical systems.

[9] Poincaré [23] first observed that the (Newtonian) 3-body problem exhibits chaotic behavior. In the $n$-body case see Miller [18], Lecar [15], Kandrup and Smith [12], and Goodman *et al* [8].

Chaotic gravitational waves would impose a constraint on any measurement, by jiggling any measurement equipment in an unpredictable way. An exact prediction in a chaotic system requires complete information about the system. However, the model for prediction is itself part of the chaotic system. The subset of the system used for prediction must incorporate a complete model of the entire system. Furthermore, to make predictions about events before they happen, the subsystem would need to prepare a forecast faster than the entire system itself evolves. This is clearly unlikely, although we do not have a formal proof. Thus, unless a relativistic *n*-body system can model itself locally, any observer that is part of that system will be constrained by an uncertainty principle.

### 3.2. Non-Locality

In light of the above, non-locality, such as violations of the Bell inequality, appears to be a more fundamental point of tension between relativity and quantum mechanics. An *n*-body relativistic system is subject to the speed of light, and an observer capable of predicting the system's behavior will not perceive any Bell violations except by coincidence or conspiracy.[12] Nonetheless, a better understanding of how chaotic gravitational waves interact with the measurement equipment of an observer subject to the above uncertainty principle would help delineate which quantum phenomena might have gravitational explanations.

---

[10] Stochastic gravitational background radiation is a focus of theoretical and empirical work, as reviewed in Maggiore [17].

[11] Lasota and Mackey [13] describe how chaotic systems can produce phenomena that appear stochastic.

---

[12] See however Hadley [9].